\documentclass[journal=JPC, manuscript=article]{achemso}

 \usepackage{graphicx,bm,amssymb,amsmath}
 \usepackage{color}
 \usepackage[utf8]{inputenc}
 \usepackage[T1]{fontenc}
 \usepackage{mathptmx}
 \usepackage{soul}

\newcommand{\codeline}[1]{\texttt{\detokenize{#1}}}

\author{M\'arcio S. Gomes-Filho}
\affiliation[universidade1]{Centro de Ci\^encias Naturais e Humanas, Universidade Federal do ABC, Santo Andr\'e, 09210-580, S\~ao Paulo, Brazil}
\email{marcio.sampaio@ufabc.edu.br}

\author{Alberto Torres}
\affiliation[universidade2]{Instituto de Física, Universidade de São Paulo,  São Paulo, 05508-090, Brazil}

\author{Alexandre Reily Rocha}
\affiliation[universidade3]{Institute of Theoretical Physics, São Paulo State University, Campus S\~ao Paulo,  01140-070, Brazil}

\author{Luana S. Pedroza}
\affiliation[universidade1]{Centro de Ci\^encias Naturais e Humanas, Universidade Federal do ABC, Santo Andr\'e, 09210-580, S\~ao Paulo, Brazil}
\email{l.pedroza@ufabc.edu.br}

\title{Size and Quality of Quantum Mechanical Data Sets for  Training Neural Network Force Fields for Liquid Water}

\begin{document}

\begin{abstract}
Molecular dynamics simulations have been  used in different scientific fields to investigate a broad range of physical systems.  
However, the accuracy of calculation is based on the model considered to describe the atomic interactions. In particular, \textit{ab initio} molecular dynamics (AIMD) has the accuracy of density functional theory (DFT), and thus is limited to small systems and relatively short simulation time. In this scenario, Neural Network Force Fields (NNFF) have an important role, since it provides a way to circumvent these caveats. In this work we investigate NNFF designed at the level of DFT to describe liquid water, focusing on the size and quality of the training data-set considered. 
We show that structural properties are less dependent on the size of the training data-set compared to dynamical ones (such as the diffusion coefficient), and a good sampling (selecting data reference for training process) can lead to a small sample with good precision.

\end{abstract}
\section{Introduction \label{sec:intro}}

Molecular dynamics simulations have been  used in different scientific fields to investigate a broad range of physical systems, such as thermodynamic properties of liquids, physico-chemical aspects of interfaces and  biomolecules.\cite{md01, md02, md03}
Its success relies on a number of  factors, for example,  the functional form assigned to describe inter- and intra-atomic interactions, the parametrization  procedure (obtaining the potential parameters), and the quality of data employed - experimental or \textit{ab initio} one.\cite{Batzner2022, Zhang18, Wen22, Zhang19, Zhang18, DeepMD18, Behler11, Pedroza08}  
Most of the classical potentials are physically and/or chemically motivated, in which a simple analytical functional form is usually considered such as the Lennard-Jones potential.\cite{Jones1924} 
As a consequence, transferability, and accuracy are a common issue in this area of  research.\cite{Zhang18, Zhang19, Unke21} 

In the particular case of water, many classical empirical models have been proposed to describe its properties. Although some classical models, such as the MB-Pol, provide fairly good results for water\cite{mb-pol_01,mb-pol_02,mb-pol_03}, there is no single water model capable of exactly reproducing all experimental results~\cite{Kadaoluwa2021}.
In fact, over the last decades, there has been an advance in the understanding of the properties of water both by theory/simulations as experimentally. However, there are still some of its properties which are not yet fully understood. For example, the microscopic origin of the water anomalies.\cite{Brini2017}
From a microscopic point of view, the quantum nature of the hydrogen bond network, the interplay between short and long range interactions and nuclear quantum effects make the liquid water intrinsically difficult to be modelled.
In this way, based on the nature of the phenomena that governs the physical and chemical properties of liquid water, first principles simulations seem to be the most appropriate choice, since they have, by construction, an accurate predictive potential. These types of simulations have the advantage of forgoing the requirement for a model or the parametrization of any experimental data. In particular, \textit{ab initio} molecular dynamics (AIMD) allows one to obtain the energy/forces on-the-fly by a quantum mechanical method such as density functional theory (DFT)\cite{Car85} at each time step. The caveat is the limitation to small system sizes and short simulations time.\cite{Zhang18} Furthermore, the quality of the AIMD simulations is closely determined by the chosen exchange-correlation (xc) functional.\cite{Gillan16,Lin2009,Ruiz2018}

In this scenario, machine learning (ML) potentials - also known as ML force fields\cite{Unke21} - have introduced a paradigm change as one can now combine the quantum accuracy of AIMD with computational efficiency of  empirical interatomic models.
This allows one to simulate large systems for long time with  \textit{ab initio} accuracy.
These methods have been recognized as promising alternatives to underline 
new physical phenomena and aid in materials-discovery processes.\cite{Waters22, Wen22, Unke21, Friederich21, Kocer22, Behler2017, Torres21,Deringer19, Noe2020, Ceriotti21, Rupp12} In particular, the microscopic comprehension of bulk water can benefit from computer simulations based on ML potentials.\cite{Torres21, Zhang21, Monserra20,Reinhardt2021, Schra21} 

Many different ML methods  have been used to construct these ML-based potentials, for example artificial neural networks,\cite{Blank95,Behler07,Schutt18, Lorenz2004, PANNA,Batzner2022, DeepMD18, Wen22, Gao22} kernel-based methods,\cite{Rupp12, Chmiela19} gaussian approximation potentials,\cite{Bartok14, Bartok15} and atomic cluster expansion.\cite{Drautz19}
In particular, deep Neural Networks (NN) approaches have been shown to be a versatile tool able to produce accurate Force Fields (FF)  trained with DFT calculations.\cite{Zhang19, Torres21, Wen22}   
The successful/accuracy of  ML potentials is directly related to the quality and size of the training data-set employed.\cite{schleder2019dft,Schleder20a, Unke21} 
Usually, deep NN approaches require a large amount of data, but tipically provide a high accuracy.\cite{Butler2018,Friederich21, Batzner2022} 
Therefore,  having a  deep NN potential trained  with less DFT data reference is a very important issue\cite{Imbalzano2018, Batzner2022}, since the DFT calculation of the reference data-set is highly computational demanding.

In this work we investigate deep NNFF designed at the level of DFT to describe liquid water, focusing on the size and quality of the training data-set considered. 
Here, we chose to investigate the \textit{ab initio} training data-set  based on {the} SCAN functional\cite{Sun15}, since it has shown some promising results for water.\cite{Chen17, Zheng18, LaCount19, Yao20, GartnerIII20, Piaggi21, Zhang21, malosso22,Torres21}. 
We show that correctly sampling the data (selecting reference data for the training process) is a crucial step, and devising a method to efficiently obtain uncorrelated structures that provide a good distribution over the phase space allows one to significantly reduce the amount of data and the size of the NN required to have accurate NNFF. As a result we also show that the structural properties are less dependent on the size of the training data-set compared to dynamical ones (e.g, {the} diffusion coefficient.)

\section{Methods\label{sec:met}}

\subsection{Computational Details\label{sec:comp}}

A crucial step in the development of the NN force field was to carefully select the bulk water configurations, which included {configurations with} long and short OH bonds before comput{ing} the DFT energies and forces used in the training process.
The protocol to obtain those configuration was: $(i)$ the selected configurations were obtained considering nuclear quantum effects (NQE) in classical MD simulations by carrying out partially adiabatic centroid molecular dynamics (PACMD)\cite{Hone06} simulations using a flexible water model (q-TIP4P/F force field\cite{Habershon09}) for a system composed by 64 water molecules; $(ii)$ good phase space sampling was obtained by performing simulations with different temperatures (T = 300 and 600 K) and densities ($\rho$ = 0.88, 1.0, and 1.2 g/cm$^3$), and $(iii)$  selecting uncorrelated configurations (i.e, geometrical structures) through radial distribution functions, choosing those that maximized the Jensen-Shannon distance.\cite{2020SciPy}
In this way, a broad range of intra- and intermolecular geometric configurations were present in the training set.\cite{Torres21} 

For each set of PACMD simulation (different T and $\rho$), the geometries were collected every 100 fs from a total simulation time of 1 ns ($10^4$ configurations). After geometric selection criteria, the number of configurations were 5000 (1000) for T = 300 K (600 K), which resulted in $18\times10^3$ snapshots. Then, for each configuration we performed a single point DFT calculation to obtain the total energies and forces. In {our} particular {case}, the reference data were obtained using the Vienna \textit{Ab initio} Simulation Package (VASP)\cite{vasp} and SCAN functional\cite{Sun15}.
The plane wave basis was set up to an energy cutoff of 1600 eV (118 Ry), and the core-valence interaction was treated by the projected augmented wave (PAW) method.\cite{PAW} {Note that all DFT calculations are completely independent from each other, and thus can be performed separately.}

\subsubsection{Training process}

A crucial point in the design of neural network-trained force fields is to determine the minimum size of the training set to obtain \textit{ab initio} quality results. In this way, we trained our NNFFs using a randomly selected subset of the training data set ranging from 10\% to 100\% of the total data set.

We first selected 90\% of the configurations as the training set and 10\% as the testing set for assessment, which means 16200 (training) e 1800 (testing) structures.
It is important to emphasize that the data used in the training is  $n_{data} = n_{energy} + n_{force}$. We have one energy value per configuration (64 H$_2$O), $n_{energy} = 16200$. Whereas for force, there are three components, \{f$_x$, f$_y$, f$_z$\}, for each atom, i.e $n_{force} = 16200 \times 3 \times 192$. Therefore, we have $n_{data}=9347400$ training data. 

On the other  hand, in order to avoid over-fitting  (when the model performs worse on the testing data than on their training set) the number of  fitted parameters, $n_{parm}$, cannot be {larger} than {the} number of training data points.
In this work, we chose $n_{data}/n_{parm}$ to be at least 8.  Note that in order to avoid over-fitting one can either increase the size of the training {set} or reduce the number of layers and/or the number of neurons of each layer.\cite{DeepMD18, chollet2021deep, Wen22}

Therefore, we scaled down the neural network in order to keep the total number of fitted parameters in the same proportion. For example,  the NN topology (the number of hidden layers is 4 and the number of neurons in each layer) is set to (32, 16, 4, 2) and (320, 160, 32, 16)  for the 10\% and 100\% case, respectively.

We used the current version of {the} DeePMD-kit code\cite{DeepMD18} to generate deep neural network potentials for bulk water based on {the} SCAN functional. In particular, we use the Deep Potential-Smooth Edition descriptor, where the full  relative coordinates are used to build the descriptor.\cite{Zhang18b, Wen22} The number of hidden layers is kept fixed, and the hyperbolic tangent was used as an activation function in the hidden layers.  The loss function was minimized with the Adam stochastic gradient descent method,\cite{adam} composed by the mean squared errors of the energies and forces with  a starting  and stopping learning rate equal to $10^{-3}$ and $3.51\times10^{-8}$, respectively. The training process undergoes $2\times10^{6}$ steps in total. Further computational details can be found in the Supplementary Information~(SI). 

\subsubsection{Deep NN Molecular Dynamics}

After we have obtained different deep NNFF for liquid water trained with subsets of our DFT training data-set, we can then perform deep NN Molecular Dynamics (NN MD) using the LAMMPS simulation package~\cite{Lammps22} and the DeepMD plugin.\cite{DeepMD18} In this way, we can investigate the effects of the training data set size on the convergence of physical properties of water.

Simulations of water at different temperatures $T$ and pressures $P$ were performed  to investigate the density convergence as a function of the training set size. These systems were first equilibrated over 50 ps by performing an isothermal-isobaric (NPT) simulations (using Nose-Hoover thermostat and barostat~\cite{Lammps22}). The equilibrium densities were then obtained averaging over 2~ns. 

We also carried out NVT simulations of  bulk water (512 molecules), controlling the temperature via a stochastic velocity rescaling thermostat\cite{thermostat}. These large systems were equilibrated over 150 ps and then additionally 2 ns simulations were carried out at the production stage.

\section{Results and discussion \label{sec:results}}

\begin{figure}[!htbp]
    \centering
    \includegraphics[width=1\columnwidth]{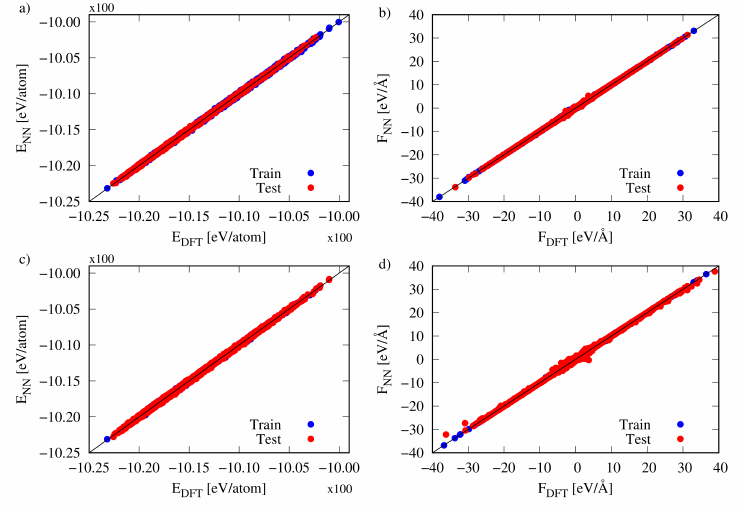}
    \caption{Parity graphs for energy and forces evaluated on the training and test data sets, where a) and b) refer to the 100\% case while c) and b) to the case of 10\%. 
    The E$_{NN}$ and F$_{NN}$ are the NN predicted energy and force, respectively. The DFT data reference are represented  by E$_{DFT}$ and and F$_{DFT}$.}
    \label{fig:parity100}
\end{figure}

In order to illustrate the performance of our deep NN potential, we show in Figure~\ref{fig:parity100} the parity graphs for the (a)  energy and (b) force components. In this particular case, we tested our model on the test set (10\% of DFT data reference) and also on the training set {itself} (90\% of DFT data reference), which shows that the NN operates well on both data sets with roughly similar {errors}, that is a good feature to indicate that the fitted NN is neither under-fitting nor over-fitting.\cite{DeepMD18}

In particular, we find that the RMSE{s} (Root Mean Squared Error{s}) on the  test set are $\sim46$ meV/\AA~ (force) and  $\sim0.53$ meV/atom (energy).  As recently pointed out by Wen and collaborators\cite{Wen22} the RMSE for forces and energy for a good  (high accuracy) deep potential should be of the order of $<50$ meV/\AA~ and $\sim1$~meV/atom, respectively.

We performed a  k-fold cross-validation procedure (see SI for technical details), where our DFT reference data (18,000 {snapshots}) were separated into 10 subsets of equal size. We then trained/tested different models with different training subsets, while keeping the proportion of 90\% for training and 10\% for testing. We show in SI the RMSEs obtained (Table I- SI).
We find that the testing errors do not change {considerably}  with different training subsets, and the average of RMSE errors for energy and forces are equal to 0.533 meV/atom and 
46.6 meV/\AA~, respectively. The deviations are in the order of 0.002\% (energy) and 0.12\% (force), which represents that the DFT reference data are uncorrelated.

\begin{figure}[!htbp]
    \centering
    \includegraphics[width=1\columnwidth]{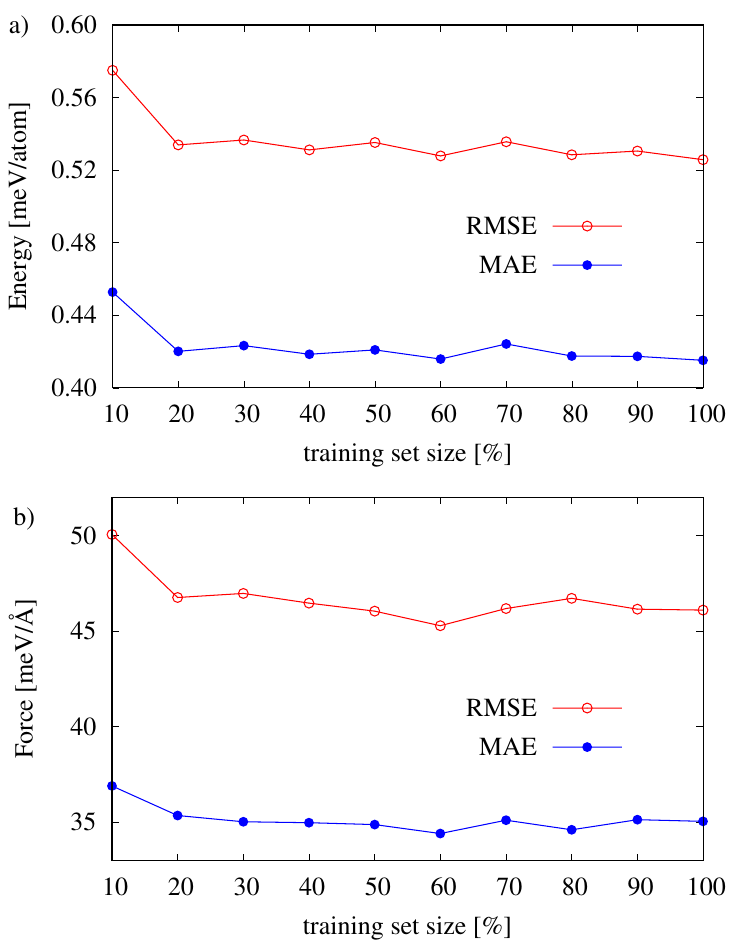}
    \caption{RMSE and MAE for energy (top panel) and forces (bottom panel) as a function of the training set size.}
    \label{fig:errors}
\end{figure}

In other words, the cross-validation tell us that the trained NNFF does not depend on a particular partition of the DFT data-set. Thus, we can train different NN using a randomly selected subset of the training data set. To exemplify it, we show in Figure~\ref{fig:parity100} c) and d) the parity graphs for the 10\% case (1620 randomly selected  snapshots used to train a deep NNFF), which shows that this NNFF  also performs well on both the testing and training data-sets.
In this way,  we show  in Figure~\ref{fig:errors} the RMSE and MAE (Mean Absolute Error) for energy (top panel) and force (bottom panel) as a function of the training data set size.

As we can see  in Figure~\ref{fig:errors}, the  four curves show an initial decrease in error values, after the size of the data-set has reached 20\%, they become essentially constant. However, it should be mentioned that even the 10\% case is within the accuracy reported in the literature.\cite{Wen22} 

Nonetheless, the observation only based on the errors evaluated on test data sets are not enough to affirm that the NNFF will work for a long MD simulation.  For example,  we considered a different case, where the NNFF was built with correlated data obtained from a 330 K NVT AIMD simulation with van der Waals exchange–correlation functional (vdW-BH\cite{vdW-BH}).  In this case, the training data set has a size similar to the 50\%  uncorrelated case presented in Figure~\ref{fig:errors}.
Although we found the RMSE errors on test data-set smaller than those presented in Figure~\ref{fig:errors},
we can not simulate bulk water for a long  time;  the energy is only conserved up to $\sim$20 ps, after that the system makes non-physical bonds between the atoms and the energy is no longer conserved. See SI  for further details.

On the other hand, the deep NNFF trained with only uncorrelated 1620 frames (10\% case) allows one to simulate water systems for long times, which shows  that  correct phase sampling  is more important than just the amount of data  used in the training process. 
For example in Table~\ref{tab:densities},  we show the  equilibrium water densities, $\rho_{eq}$, obtained at different temperatures and at fixed ambient pressure  for the 10\% and 100\% cases. We also show the result for the hexagonal ice I$_{\text{h}}$ with 96 H$_2$O at $T=273$~K and $P=1$~Bar.

 \begin{table}[!htbp]
  \resizebox{7cm}{!}{
     \begin{tabular}{c c c}
        \hline
        \hline
                                                        &  10\%                         & 100\%                  \\     
                                                        &     $\rho_{eq}$ [g/cm$^3$]    & $\rho_{eq}$ [g/cm$^3$] \\
    \hline
        water: T = 300K    & 1.015                          &     1.013   \\
        \hline
        water: T = 350K    & 1.020                          &     1.024  \\
            \hline
     ice I$_{\text{h}}$: T = 273K      & 0.957                           &     0.961  \\
        \hline
         \hline
    \end{tabular}
 }
    \caption{\label{tab:densities}  Equilibrium densities, $\rho_{eq}$, for liquid water with 128 H$_2$O at pressure equal to 1 Bar and temperature equal to 300 K and 350 K. We also show the density result for ice I$_{\text{h}}$ with 96 H$_2$O at $T=273$~K and $P=1$~Bar. }
 \end{table}

It is worth mentioning that for liquid water both NNFFs result in densities with the same precision although one was trained with 1620 samples and another one with ten times more data points. Moreover, it also captures the ice I$_{\text{h}}$ density in  reasonable agreement with other SCAN results under similar thermodynamic conditions (AIMD:\cite{Chen17} 0.964$\pm$0.023; SCAN DFT\cite{Piaggi21} 0.957$\pm$0.004; NNFF\cite{Piaggi21} 0.949$\pm$0.001).
 
Another interesting point is that although our DFT data reference was obtained for liquid water at different temperatures and densities, the NNFF were able to estimate the ice density.  In fact, the capability of a ML potential trained on liquid water alone  predicting the  properties of the ice phases was also  recently reported by Monserrat et al.\cite{Monserra20}

\begin{figure}[!htbp]
    \centering
    \includegraphics[width=.82\columnwidth]{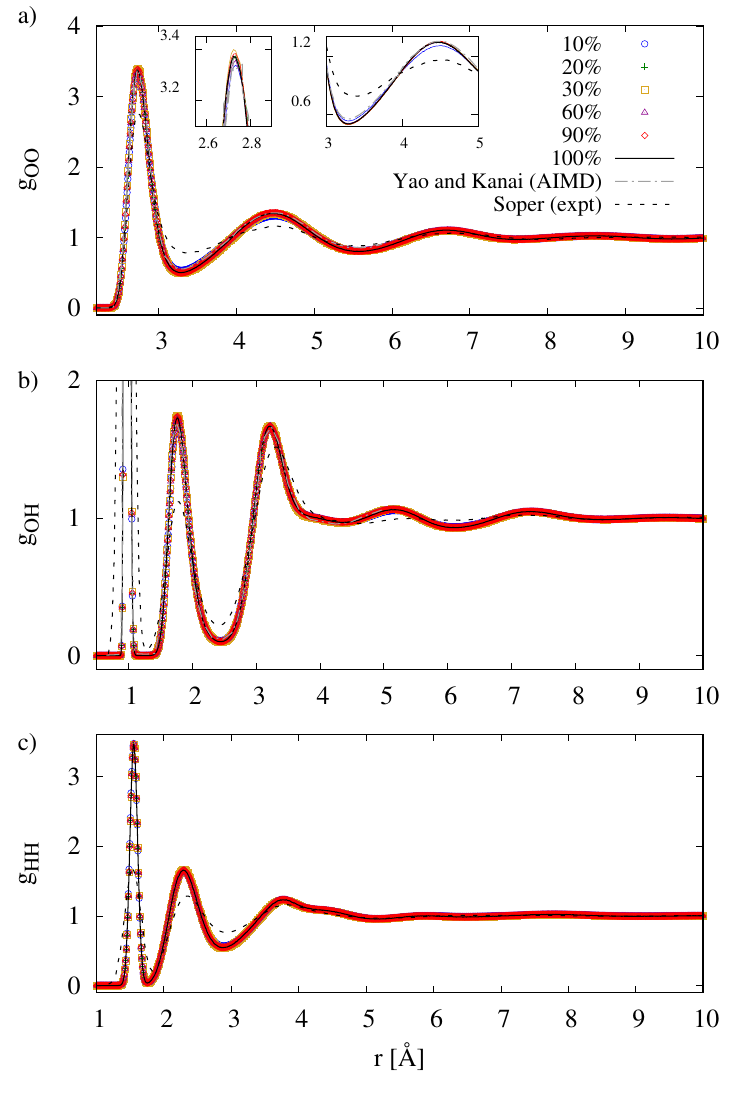}
    \caption{ Radial distribution function for (a) O-O, (b) O-H and (c) H-H pairs obtained via NNMD. Each percentage presented refers to the fraction of the  original data set used for training. The g$_{OO}$ and g$_{OH}$ results\cite{Yao20} obtained from AIMD-based on SCAN functional for 55 water molecules at T = 300 K are represented by dashed lines. The insets in g$_{OO}$ panel show the points of the first and second peaks. The experimental results from Soper\cite{Soper20} are also shown.}
    \label{fig:rdfs}
\end{figure}

In  Figure~\ref{fig:rdfs} we show the pair correlation functions for oxygen–oxygen (g$_{OO}$), oxygen–hydrogen (g$_{OH}$) and hydrogen–hydrogen  (g$_{HH}$), which are  the main structural descriptors for water.\cite{SOPER1998, Soper20, Vega05} These results were obtained for liquid water composed by 512 molecules at fixed density ($\rho = 0.997$ g/cm$^3$) and temperature (T = 300 K). 
We also show the  AIMD results for the SCAN functional for 55 water molecules at the same temperature recently presented by Yao and Kanai,\cite{Yao20}. As it can be seen, all results are very similar.

\begin{figure}[!htbp]
    \centering
    \includegraphics[width=1\columnwidth]{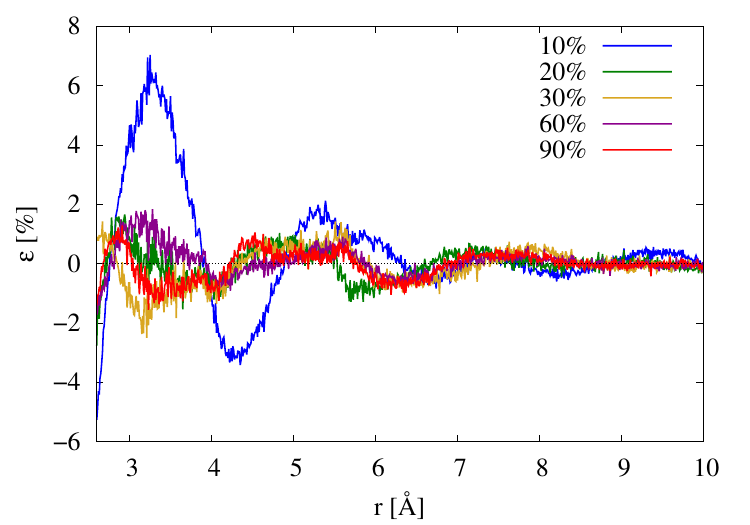}
    \caption{
    The relative error, of the O-O radial distribution function for different data set sizes, as a function of the distance, $r$, in angstrom.}
    \label{fig:error_rdf}
\end{figure}

For a better comparison between the NNFF, we measured the relative error between the radial oxygen-oxygen distribution function for each trained NNFF  with respect to the hundred percent case, as it is shown in Figure~\ref{fig:error_rdf}.
Note that the errors are roughly the same, fluctuating around zero ({approximately} 2\% of error). The only  exception is  the 10\% case that presents errors of    $\sim6$\% in the range of $3 < r < 5$~\AA.

Furthermore, we also analyzed the vibrational spectra obtained from different molecular dynamics simulations (deep NNFF trained with 10\% and 100\% of the training data-set). All of them have approximately the same pattern, which means that the  results for the structural and vibrational properties of bulk water are mostly independent of the size of the training data-set employed to built deep NNFF for liquid water, as long as they have been well chosen. 

\begin{figure}[!htbp]
    \centering
    \includegraphics[width=1\columnwidth]{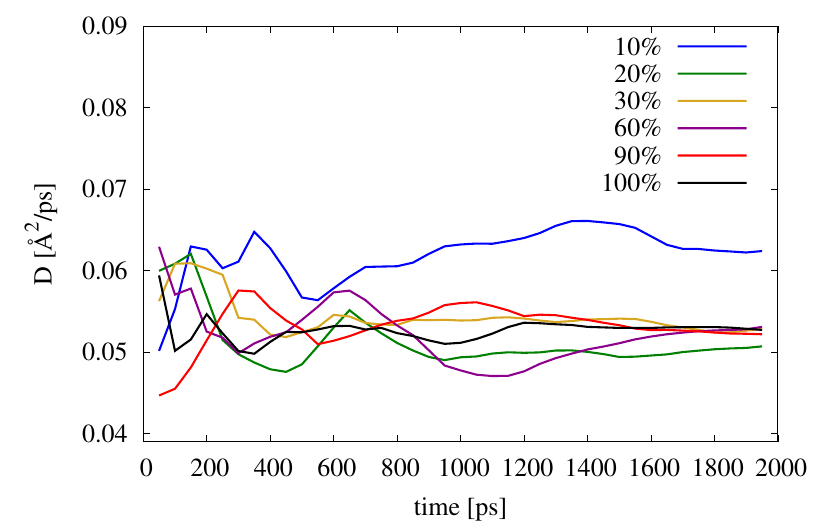}
    \caption{Self-diffusion coefficient as a function of simulation time, for 512 water molecules at T = 300 K and $\rho = 0.997$ g/cm$^3$.}
    \label{fig:diffusion}
\end{figure}

More importantly, since MD simulations can now be performed for a long time scale, we can investigate the dynamical properties as a function of time. An important feature of bulk water that is not usually well described by AIMD is its self-diffusion coefficient.\cite{Torres21} In fact, the self-diffusion coefficient can depend on many factors even when obtained from a classical MD simulation.\cite{Tsimpanogiannis19, Kadaoluwa2021}
In Figure~\ref{fig:diffusion} we show   the self-diffusion coefficient as a function of simulation time, where it  was obtained from the Einstein equation of the mean square displacements
of the center of mass of water molecules. For further technical details see Ref. \cite{Torres21}.
As it can be seen, as the time increases the fluctuations are reduced and the value of $D$ converges. As already shown, this only occurs after 2 ns, way above conventional AIMD capabilities.  
At a final time (2000 ps),  the 10\% case is the only one that presents a higher deviation with respect to the 100\%.

It should be mentioned that in earlier works,\cite{DeepMD18, Zhang18} the training data-set used to build the ML force fields typically came from AIMD simulations. For example, the data obtained for short simulation time ($\sim$20 ps), result{ed} in a total of 40$\times10^3$ correlated configurations.\cite{DeepMD18} On the other hand, {more recently} active learning procedure{s} {have} allow{ed} {for the construction of} a ML potential with fewer training data points.\cite{Zhang19} {In the work of Malosso {\it et al.}, f}or example, the ML potential for liquid water {was} trained with 4000 configurations,\cite{malosso22} which is in agreement with the size of the training data-set required for the convergence of the self-diffusion coefficient  present in Figure~\ref{fig:diffusion}. Thus, it is possible that combining uncorrelated snapshots and active learning the size of the data set could be reduced even further.

\section{Conclusions \label{sec:con}}

In this work we analyzed deep neural network potentials for liquid water/ice I$_{\text{h}}$ based on the SCAN xc-functional.\cite{Sun15} We show that the structural properties such as the equilibrium densities at different temperature and pressures  for water are quite independent of the size of the training data-set considered, as the minimum amount employed here was the energy and forces of 1620 structures (64 H$_2$O). This quantity is much {smaller} than {what is typically} used in the training of other NN. This can be attributed to the method of selecting the snapshots for training, which provide a breadth of structures that best samples the phase space.
In this way, we have found that the density,  vibrational spectra and  the radial distribution function of water are less dependent on the size of the training data-set compared to dynamical ones (e.g, {the} diffusion coefficient.)
{Finally, we envision}  th{at} {uncorrelated} physical inspired training data-set procedure proposed here (sets tha include a broad range of short and long OH bonds) together with active learning, can be a way to produce reliable ML potentials constructed with fewer DFT training data points.

\begin{suppinfo}
The Supplementary Information is available free of charge at xxxxx.

\begin{itemize}
  \item Supplementary information contents: training details, cross-validation, molecular dynamics results, and  correlated data.
\end{itemize}
\end{suppinfo}

\begin{acknowledgement}

We acknowledge fruitful discussions with Gabriel C. Santucci and Lucas T. S. de Miranda.
The authors acknowledge financial support from \-FAPESP\- (Grant \# \-FAPESP\- 2017/10292-0, 2020/09011-9, 2017/02317-2 and 2016/01343-7).
This work used the computational resources from the Centro Nacional de Processamento de Alto Desempenho em São Paulo (CENAPAD-SP) and GRID-UNESP.
\end{acknowledgement}

\bibliography{references}

 \begin{figure}[!htbp]
    \centering
    \includegraphics[width=1\columnwidth]{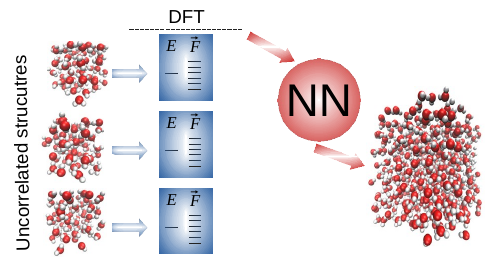}
    \caption{TOC graphic.}
    \label{fig:toc}
\end{figure}

\newpage 

\begin{center}
    \textbf{\Large Supporting Information: Size and Quality of Quantum Mechanical Data Sets for Training Neural Network Force Fields for Liquid Water}
\end{center}

\section{SI note 1: Training parameters}

We have used the current version of the package DeePMD-kit v2.1.3~\cite{SI-deepmd} to generate deep Neural Network Force Fields (NNFF) for bulk water based on the SCAN functional.\cite{SI-scan} A different deep-NNFF was obtained for each subset of the training data-set. The DFT data reference employed here were obtained in a previous work\cite{SI-torres} (see main text for details on the training data-set). 
 
In particular, we have used the Deep Potential-Smooth Edition descriptor, in which the full relative coordinates are used to build the descriptor.  
We trained our neural network (NN) using a randomly selected subset of the training data set ranging from 10\% to 100\% of 16200 samples. We scaled down the NN in order to maintain the total number of fitted parameters and avoid overfitting. For example, the NN topology was chosen to be (32, 16, 4, 2)  and (320, 160, 32, 16)  for the 10\% and 100\% case, respectively.
 
For all trained deep NNFF, we kept fixed the size of the embedding (25,50,100), the cutoff radius (6~\AA~), and the smoothing parameter was chosen to be equal to 0.50~\AA~ (similar to the NNFF proposed for water by Zhang et al.\cite{SI-zhang}).
The possible maximum number of neighbors in the cut-off radius is set to be 46 and 92 for oxygen and hydrogen atoms, respectively. 
The flag \codeline{axis_neuron} is set to 16, which means that this is the size of submatrix of the embedding matrix, for details see.\cite{SI-smooth}
 
The loss function is composed by the mean squared errors of the energies and forces, where its hyper-parameters were chosen to \codeline{start_pref_e =	0.02}, \codeline{limit_pref_e = 8}, \codeline{start_pref_f = 1500} and \codeline{limit_pref_f = 1}.

The loss function was minimized with the Adam stochastic gradient descent method until $2\times10^6$ steps with the starting and stopping learning equal to $10^{-3}$ and $3.51\times10^{-8}$, respectively.

\section{SI note 2: cross-validation}

\begin{figure}[!htbp]
    \centering
    \includegraphics[width=0.8\columnwidth]{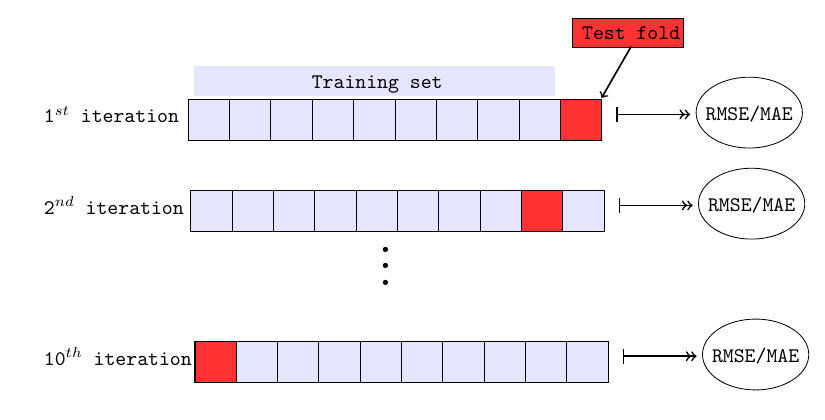}
    \caption{K-fold cross-validation diagram, where $K=10$, i.e 90\% of the DFT data reference is used for training process, while the reminiscent 10\% is used for testing. The error (RMSE/MAE) is obtained for each iteration and the final average results are listed in Table~\ref{tabSI:cross}. }
    \label{figSI:kfolcross}
\end{figure}

We performed a K-fold cross-validation procedure in order to verify whether the DFT reference data~\cite{SI-torres} are correlated or not. The K-fold cross-validation essentially consists in partitioning the data set into $K$ equal parts, then $K-1$ partitions are used for training and one is designated for testing,\cite{SI-Chollet, SI-Schleder} as it is illustrated in Figure~\ref{figSI:kfolcross}. Note that this process is performed K times.

This cross-checking is usually assigned to test the machine learning approaches in situations where the number of data sets is limited.\cite{SI-Schleder} In our case, we used it to verify if the deep neural network potential depends on a particular partition of the DFT data set considered. 
With respect to NN accuracy, we computed the errors values for both energy and force by MAE (Mean Absolute Error):
\begin{equation}
    MAE = \frac{1}{n} \sum_i^n | y_i - \hat{y_i}|,
\end{equation}
and by RMSE (Root Mean Squared Error):
\begin{equation}
    RMSE = \sqrt{\frac{1}{n} \sum_i^n ( y_i - \hat{y_i})^2},
\end{equation}
where $n$ is the number of samples, $\hat{y_i}$ the DFT reference (energy or force) and $y_i$ the predicted value.

  \begin{table}[!htbp]
 \resizebox{14.cm}{!}{
     \begin{tabular}{c c c c c}

                                                &    Energies [eV/atom]       &  &    Forces [eV/\AA]     \\
            \hline
           k                           & RMSE & MAE    &     RMSE & MAE        \\     
            \hline
1   &    0.000526 &  0.000415 & 0.046102 &  0.035042 \\ 
2   &    0.000524 &  0.000414 & 0.045933 &  0.035076  \\ 
3   &    0.000544 &  0.000431 & 0.047203 &  0.036042  \\ 
4   &    0.000552 &  0.000435 & 0.048040 &  0.036647  \\ 
5   &    0.000516 &  0.000408 & 0.045855 &  0.035020  \\ 
6   &    0.000543 &  0.000426 & 0.047148 &  0.036018  \\ 
7   &    0.000516 &  0.000406 & 0.045123 &  0.034423  \\ 
8   &    0.000522 &  0.000411 & 0.045615 &  0.034856  \\
9   &    0.000567 &  0.000447 & 0.049275 &  0.037491   \\ 
10  &    0.000522 &  0.000411 & 0.045811 &  0.035023   \\ 
\hline
average &   0.000533 &  0.000420 & 0.046611 &  0.035564 \\
 std &   0.000016 &  0.000013 & 0.001221 &  0.000910 \\
\hline
    \end{tabular}
 }
 \caption{\label{tabSI:cross}: 10-fold cross-validation: values of the RMSE and MAE on test data-set.  All NNs present similar results. The last two lines designate the average and the standard deviation (std) values.}
 \end{table}

We show the RMSE and MAE obtained for each iteration of the 10-fold cross-validation procedure in Table~\ref{tabSI:cross}, where the last two lines represent the average and its standard deviations (std). Note that each standard deviation is quite small, meaning that testing errors do not change considerably with different training subsets, which means that the DFT reference data are uncorrelated.

\section{SI note 3: molecular dynamics results}

\subsection{Structural properties}

\begin{figure}[!htbp]
    \centering
    \includegraphics[width=0.8\columnwidth]{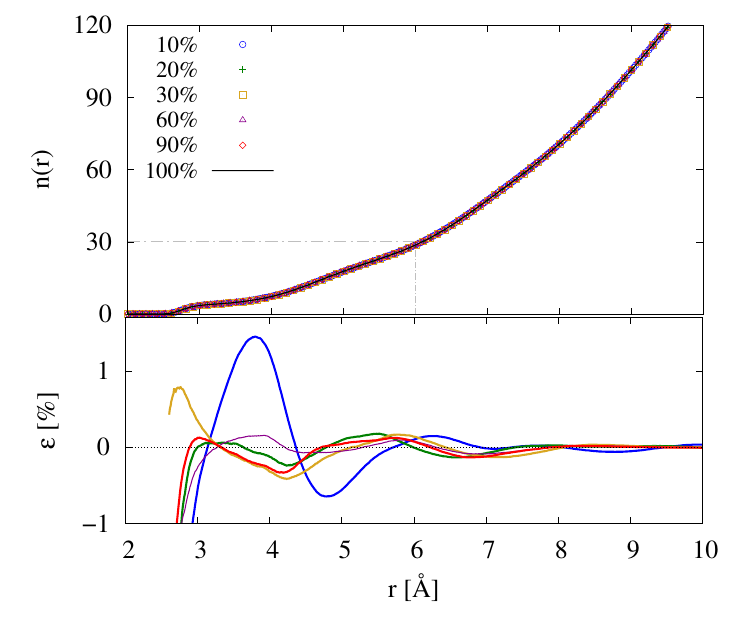}
    \caption{Top to bottom: integration number, $n(r)$, as a function of the distance $r$ in angstrom for the oxygen-oxygen pair; integration of $g_{OO}(r)$. The relative error, $\varepsilon[{\%}]$, with respect to the hundred percent case. Each percentage is the quantity of data points used in the training process, for instance, 10\% is equivalent to 1620 samples. }
    \label{figSI:rdf}
\end{figure}

The results here are for bulk water simulation presented in the main text for a system composed by 512 H$_2$O molecules at 300 K.
We calculated $n(r)$, which is the integral of the radial oxygen-oxygen distribution function, $g_{OO}(r)$. This represent  the number of oxygen atoms {at} a given distance $r$ (coordination number) (shown in Fig.~\ref{figSI:rdf}). 

We also show the relative error, $\varepsilon$,  with respect to a hundred percent case, $\varepsilon = \left  (\frac{n(r)^{\xi} - n(r)^{100}}{n(r)^{100}} \right)\times 100$,  with $\xi = \{10, 20, 30, 60, 80\}\%$. As we can see, $n(r)$ presents a similar behavior for all cases. Furthermore,  the vertical dashed line shows that the expected number of water molecules within a sphere of radius equal to 6~\AA~ was correctly obtained for all deep NNFF. 
The relative error shows that the errors are roughly the same, fluctuating around zero, exceptions only for the 10\% case in the range of $3 < r < 5$~\AA. Therefore, the results for the structural properties of bulk water are quite independent on the size of the training data set.

\subsection{Vibrational spectra}

\begin{figure}[!htbp]
    \centering
    \includegraphics[width=0.8\columnwidth]{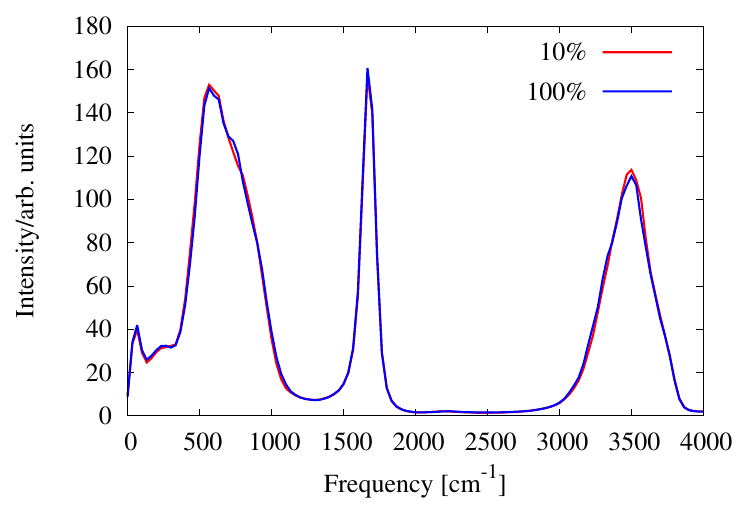}
    \caption{ The vibrational spectra of liquid water (512 H$_2$O at T = 300K) obtained by the Fourier transform of velocity autocorrelation function, equation~(\ref{eq:freq}).}
    \label{figSI:spectra}
\end{figure}

We show in Figure~\ref{figSI:spectra} the vibrational spectra of liquid water, which was obtained by the Fourier transform of the velocity autocorrelation function~\cite{SI-bookMD}, i.e,
\begin{equation}
    I(\omega) = \frac{1}{2\pi} \int_{-\infty}^\infty \langle \vec{v}_i(t)\cdot\vec{v}_i(0) \rangle  \exp(i \omega t) dt, \label{eq:freq}
\end{equation}
where $\vec{v}_i(t)$ is the velocity of the atom $i$ at time $t$ and $\omega$ is the frequency. As it can be seen, the vibrational spectra obtained from different molecular dynamics simulations (deep NNFF trained with 10\% and 100\% of the training data-set) have approximately the same pattern, which means that the  results of the vibrational properties of bulk water are quite independent of the size of the training data-set employed to built deep NNFF for liquid water.

\section{SI note 4: correlated data}

In order to show that a good sampling (selecting reference data for the training process) is a crucial step to obtain a deep NNFF, 
we have built a NNFF with the data used for training  extracted from a 330 K NVT AIMD simulation with van der Waals exchange–correlation functional, as proposed by Berland and Hyldgaard.\cite{SI-BH}  The data set was generated by simulating  a bulk liquid water system of 64 molecules using the SIESTA package.\cite{SI-siesta}  The time step was 0.50 fs and the total number of steps set up to 10$^4$. 

For training process, we discarded the first 500 frames and the 9500 frames was randomly shuffled. We then selected 90\% of the data-set for training while 10\% was used as testing data. Note that this training data set has size similar to our 50\%  uncorrelated case presented in the main text. Thus, we adopted the same training parameters as used in the 50\% case, for example,   the number of hidden layers is 4, the number of neurons in each layer is set to (160, 80, 16, 8).  

\begin{figure}[!htbp]
    \centering
    \includegraphics[width=0.8\columnwidth]{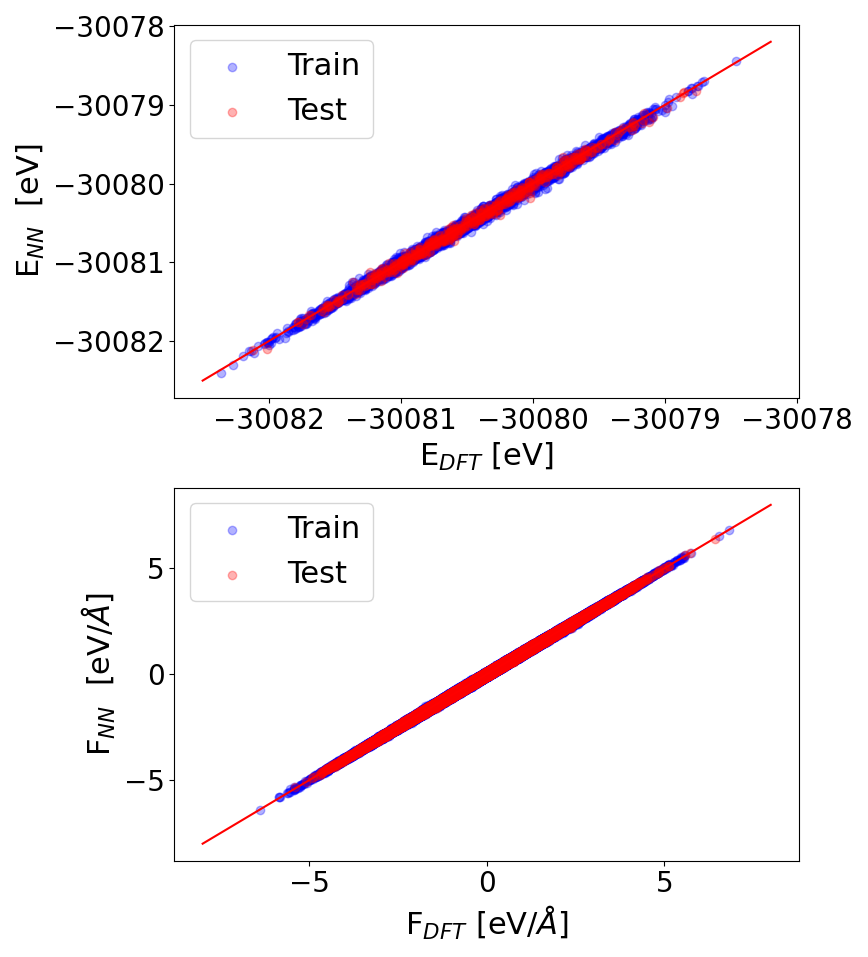}
    \caption{ Parity graphs for energy and forces evaluated on the training and testing data sets. The E$_{NN}$ and F$_{NN}$ are the NN predicted energy
and force, respectively. Already the DFT data reference are represented by E$_{DFT}$ and and F$_{DFT}$.}
    \label{figSI:corr}
\end{figure}

In figure~\ref{figSI:corr}, we show that the deep NNFF present good accuracy on both training and test data sets. The RMSE error values are listed in table~\ref{tabSI:corr}. 
Also note that these errors are smaller than those obtained from the 50\% uncorrelated case (see Fig 2 in the main text).

\begin{figure}[!htbp]
    \centering
    \includegraphics[width=0.8\columnwidth]{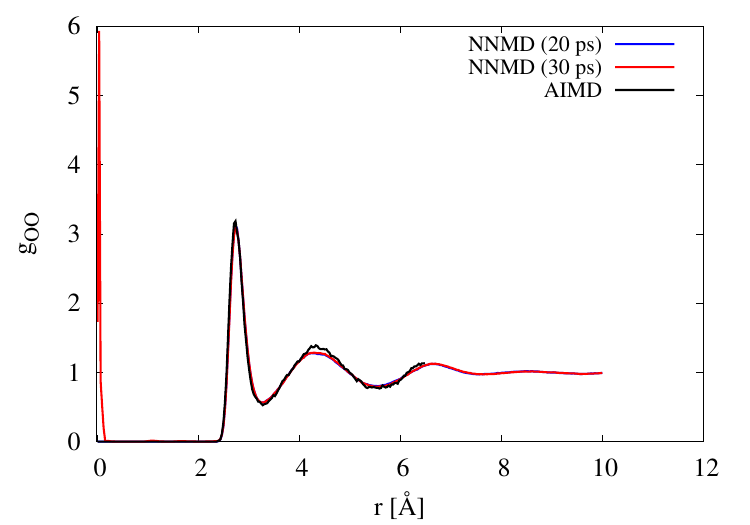}
    \caption{ Radial distribution function for O-O obtained via  AIMD (64 H$_2$O for 5 ps) and via  NNMD (512 H$_2$O) for a total of 20 ps and 30 ps. }
    \label{figSI:corr_rdf}
\end{figure}

However, the evaluation of errors on the test data set is not enough to ensure that a good NNFF for liquid water was achieved. For instance,  when we performed a simulation for a large system and for long simulation time (similar to the simulation for the bulk water presented in the main text), the energy is only conserved up to $\sim$20 ps of simulation time, after that the system makes non-physical bonds between the atoms and the energy is no longer conserved. In Figure~\ref{figSI:corr_rdf} we show the pair correlation functions for oxygen–oxygen (g$_{OO}$)  obtained via  AIMD (64 H$_2$O) and via  NNMD (512 H$_2$O) for a total of 20 ps and 30~ps.

Therefore, it should be noted that we were able to simulate  for long times the same system with a deep NNFF trained with only 1620 frames (10\% case), showing that  correct phase space sampling  is more important than just the quantity of data  used in the training process. 

  \begin{table}[!htbp]
 \resizebox{8.cm}{!}{
     \begin{tabular}{c c c }
                          & Test  & Train \\   
\hline
  Energies [meV/atom]      &  0.199   &  0.192        \\
 Forces [meV/\AA]          &  29.28   &  28.99       \\
 \hline
    \end{tabular}
 }
 \caption{\label{tabSI:corr} Neural network RMSE errors on the training and test sets.}
 \end{table}

\pagebreak

\end{document}